\def\comment#1{}
\newcommand{\be}{\begin{eqnarray}}
\newcommand{\ee}{\end{eqnarray}}

\documentstyle[prl,aps,amsfonts,psfig,multicol,epsf]{revtex}
\begin{document}
\title{{
On characteristic length scales and 
formation of vortices   in the Ginzburg-Landau-Higgs 
model \\  in the presence of a uniform background
charge
}}
\author{
  Egor Babaev
\thanks{Email: egor@teorfys.uu.se; http://www.teorfys.uu.se/people/egor/
 Tel: +46-18-4717629, Fax +46-18-533180
}
}
\address{
Institute for Theoretical Physics, Uppsala University
Box 803, S-75108 Uppsala, Sweden
}
\maketitle
\begin{abstract}
In this brief report we consider a non-local Ginzburg-Landau-Higgs model 
in the presence of a neutralizing uniform background charge.
We show that such a system possesses vortices that 
feature  a strong radial electric field. We estimate
the basic properties of such an object and characteristic
length scales in this model.
\end{abstract}
\begin{multicols}{2}
\narrowtext

The  vortex matter is a very large and diverse branch of  theoretical
physics. This concept, 
 initially introduced in superconductivity \cite{aaa},
 was later generalized to high-energy physics,  where the 
Abrikosov vortex is known as the Nielsen-Olesen string
\cite{hind}. 
Many types of vortices  were studied in detail 
both in condensed matter and particle physics \cite{volo}. 

In this paper we describe vortices in the Ginzburg-Landau model
(known in high energy community as the Abelian Higgs model)
 in the presence of a neutralizing background charge.
The key feature of this model system is that the 
charged complex scalar field interacts with 
compensating uniform background charge that, as we show below,
gives rise to a formation exotic vortices. 
The study of this model is motivated by several examples
from condensed matter physics, such as 
the Josephson junction
arrays and the granular superconductors, description of which
usually includes Coulomb terms (for a recent work and citations 
see e.g. \cite{Doniach}).
Additional motivation is the strong-coupling superconductors 
\cite{sc} where the Cooper pairs are tightly bound and 
there are no free electrons in the vortex core, although  the strong coupling 
superconductor is a more complex system than that considered in this paper.
In this brief report we do not discuss specific applications
but restrict the discussion to the simplest Ginzburg-Landau-Higgs model
of charged condensate coupled to compensating uniform background charge.

The energy functional  of the ordinary Ginzburg-Landau-Higgs model reads
\be
E=  \int d^3 x \left[ \frac{\hbar^2}{2M}\left|\left(\nabla -i \frac{e}{c} A
\right)\Psi \right|^2-  \mu \Psi^2+\frac{U}{2}|\Psi|^4
\right]
\label{gp}
\ee
The order parameter $\Psi$ 
is characterized by the coherence length \cite{fetter}:
\be
\xi_c=\frac{\hbar}{\sqrt{2M U} \Psi_0}
\ee
where $\Psi_0$ is the average value of the modulus of the complex
scalar field.

We introduce in this model a neutralizing background charge
with charge density $-e <|\Psi|^2>  = -e\Psi_0^2$.
In a such system in a presence 
of a local inhomogeneity 
of the complex charged scalar field there appears an electric field.
The energy functional for this model is non-local:

\be
E=  \int d^3 {\bf x} \biggl\{ \frac{\hbar^2}{2M}\biggl|\biggr(\nabla -i \frac{e}{c} A\biggr)\Psi\biggr|^2-  \mu \Psi^2
+\frac{U}{2}|\Psi|^4
 +
 \nonumber \\
+ \frac{e^2}{2} \int d^3 {\bf y}
\frac{[\Psi^2({\bf  x})-\Psi_0^2] [\Psi^2({\bf  y})-\Psi_0^2] }{|{\bf x-y}|}\biggr\}
\label{gp1}
\ee
Apparently, in this system   a vortex core produces a local
charge inhomogeneity and an electric field
 \cite{comm}.
The appearance
of Coulomb force in a local inhomogeneity in this system 
 makes the question of description of vortices 
to be a complicated problem.
However as we show below, 
all the parameters of the problem can be estimated
 that allows to  describe qualitatively all the 
basic properties of the vortices in this non-local model.

Existence of a zero point of the scalar field in the center 
of the core would lead to accumulation of charge $Q$ per  length $L$
of order 
\be
\frac{Q}{L}\propto - e \Psi_0^2 \times [ \mbox{vortex  core radius}]^2
\ee
In a neutral superfluid the vortex radius is defined 
by a coherence length.  
 However since the system is  charged 
the local inhomogeneity 
makes an additional energy cost due 
to appearance of electric field. Thus  the Coulomb
force  tries to shrink the vortex core size and  
this leads to a redefinition of the length scale over which such a 
system restores homogeneity. We show below
that the Coulomb force in  such a system
determines the ``healing" length scale of the field.
One of the exotic consequences 
is the possibility of conversion of the system 
from type I to type II. 

\begin{figure}[htpb]
\vspace{.3cm}
\epsfxsize=0.5\columnwidth
\centerline{
\epsffile{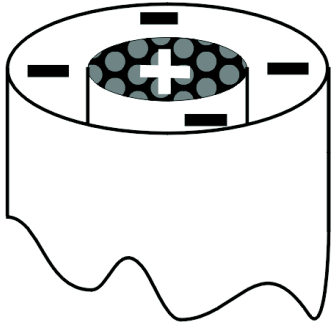}
}
\vskip 0.5cm
\caption{Structure  
of a charged vortex.  Sign ``{\bf +}" indicates the  core 
where the modulus of the complex scalar
field reaches zero 
so there exists non-compensated positive background charge.
This charge inhomogeneity induces a ``tube-like" 
area of screening charge around the vortex that is denoted by ``{\bf --}".
Apparently, the Coulomb force induced by charge inhomogeneity tends
to shrink size of the core thus leading to redefinition 
of notion of a coherence length.
 }
\label{exp}
\end{figure}

Fig. \ref{exp} 
shows a schematic picture of 
a charged vortex.
Let us estimate the ``healing" length.
The Coulomb energy of a charged vortex per length $L$
is of order of:
\be
\frac {U_C}{L} \propto e^2 \xi_{healing}^4 \Psi_0^4
\label{pl}
\ee
We can add 
to  the Ginzburg-Landau expression for the free energy
the term  
\be
e^2 \xi_{healing}^2 [\Psi({\bf x})^2-\Psi_0^2]^2,
\ee 
which estimates the order of magnitude 
for the energy of the electric field  which appears
in a local inhomogeneity.
With it we can self-consistently  determine 
the length scale $ \xi_{healing}$ 
also taking into account 
the interaction term $(U/2)\Psi^4 $.
Thus our estimate  
takes into consideration contributions both 
from the original coherence length parameter and also 
from the Coulomb force which also contributes to the  ``restoration"
of homogeneity:
\be
\xi_{healing} =\frac{\hbar}{\sqrt{2\Psi_0^2   M
\left[e^2\xi^2_{healing}+U\right]}} =\nonumber \\
\left[ -\frac{U}{2e^2} +\sqrt{
\frac{U^2}{4e^4}+\frac{\hbar^2}{2 e^2 \Psi_0^2 M}}
\right]^{1/2}
< \xi_c
\ee
When the contribution of the term $\frac{U}{2}\Psi^4$ to the 
restoration of homogeneity can be neglected, 
the healing length has  purely  Coulomb origin 
and is given by:
\be
\xi_{healing} \propto \left[ \frac{\hbar^2}{M \Psi_0^2 e^2 }\right]^{1/4} 
\ee
This can be compared with magnetic field  penetration length 
\be
\lambda=\sqrt{\frac{ M c^2}{2 \pi \Psi_0^2 e^2}}
\ee
Correspondingly for the parameter $\kappa =\lambda/\xi_{healing}$
which serves to distinguish between a system of type I ($\kappa <1/\sqrt{2}$)
and type II  ($\kappa >1/\sqrt{2}$) \footnote{for a detailed 
discussion of this property of the simplest Abelian Higgs model
in context of Nielsen-Olesen strings see e.g. \cite{hind}, an application to 
a specific superconducting systems in external magnetic fields
 requires also a study of thermodynamic
stability.} we can write:
\be
\kappa=\sqrt{\frac{M c^2}{
2 \pi \Psi_0^2 e^2 
\left[
-\frac{U}{2e^2}+
\sqrt{\frac{U^2}{4e^4}+\frac{\hbar^2}{2 e^2 \Psi_0^2 M}}
\right]}}
\ee
The above expression shows that besides $\kappa$ depends 
on coupling $U$, 
it also  depends on $\Psi_0$. It particular in the 
case when one can neglect contribution of $U$, the
 above expression  can be written as:
\be
\kappa =\sqrt{\frac{M^{3/2}c^2}{2\pi\hbar e\Psi_0 }}
\ee
Thus one of the exotic features of this system is that 
the Coulomb interaction increases the value of the parameter $\kappa$
and in principle may convert a system
from type I to type II
 when one varies the charge density $e\Psi_0^2$.

We  can also estimate the energy  per unit length 
of a vortex in this system.
Similar to  Abrikosov vortex there is a contribution form 
the magnetic field energy which density in the vortex core
is $H^2/8\pi$ and  the
kinetic energy of the supercurrent $j$:
\be
E_{kin}=\frac{M j^2}{2\Psi_0^2e^2}=\frac{\lambda^2}{8  \pi}
 (\mbox{rot}{\bf H})^2
\ee
The  difference with the corresponding discussion in the 
case of an ordinary Abrikosov vortex is that  our situation 
features also the radial electric field.
In  such a system in an applied magnetic field there is
induced in a vortex core radial electric field. The electric field
is trying to keep the core radius to be small.

In these circumstances the vortex energy per unit length  
is  given by:
\be
\varepsilon &\approx& 
\frac{1}{8 \pi} \int d^3 x[{\bf H}^2 + {\bf E^2} +\lambda^2(\mbox{rot}{\bf H})^2]
 \approx \nonumber \\
&\approx&
\frac{1}{16 \pi}\frac{\Phi_0^2  \Psi_0^2 e^2}{M c^2 } \times
\nonumber \\ 
&\times&\ln 
\left[
 \frac{M c^2}{\pi \Psi_0^2\left(-{U}+
\Psi_0M^{1/2}\sqrt{{U^2 \Psi_0^2M}+{2\hbar^2 e^2}}
\right)}
\right] + \nonumber \\
&+&  \frac{\hbar^2 \Psi_0^6}{M},
\label{nen}
\ee 
where $\Phi_0=\pi \hbar c/e$ is the quantum of magnetic flux.
\comment{
Let us also consider a more realistic
from experimental point of view case
of  moderately strong coupling strength
(or moderately low carrier density).
Then along with composite bosons there is  also a certain
fraction of non-paired carriers.  The Ginzburg-Landau
free energy equation can then be written in the form:
\be
F= -\frac{\hbar^2}{2M}(\nabla \Psi)^2-\frac{{\tilde \mu}}{2}\Psi^2+
\frac{U}{4}\Psi^4 +\frac{C}{4}|n_b-\Psi^2|^2
\label{moderate}
\ee
The last term follows immediately:
taking into account the 
Maxwell equation: ${\mbox{div}}{\bf E} = 4 \pi e[|\Psi({\bf x})|^2 -n_b ]$
and the expression for the energy of electrostatic field $U= 1/8 \pi \int E^2 d^3 {\bf x}$
we have the following additional term in the free energy functional
coming from the Coulomb interaction screened by free electrons:
\be
U\propto  e^2\int d^3{\bf x}d^3{\bf y}[\Psi({\bf x})^2-n_b]
\frac{e^{-\kappa |{\bf x-y}|}}{|{\bf x-y}|}[\Psi({\bf y}) ^2- n_b]
\ee
From this expression the last term in  (\ref{moderate})
follows immediately when the Debye radius is small.
It is  easily seen that the charging effect 
in the regime where there exist free carriers 
has the same effect of redefinition of healing length scale
of the condensate: 
\be
\xi_{healing}=\frac{\hbar}{\sqrt{2Mn_b [C+U]}}
\ \ \ \ \ \
{\mbox{(moderate coupling)}} 
\ee
}

In conclusion, we discussed the vortices formation
in the Ginzburg-Landau-Higgs model in the presence of a
compensating uniform background charge.
 We have shown that 
such a substance possesses an additional
length scale which we  call the ``healing length"
and allows  the formation of  vortices.
The key feature of the discussed vortices
is the accumulation of charge in its core and existence
of a strong radial electrostatic field. 
The study of this model is motivated by 
condensed matter models such 
as Josephson Junction arrays and granular 
superconductors
description of which 
usually includes Coulomb term \cite{Doniach,comm}.
Besides that, the above described  properties and 
length scales of the model with Coulumb interaction 
may have applications in many discussions in  
high energy physics-related models
and
are relevant for the recent 
discussion of vortex solutions in 
models like considered in \cite{antti}. 

We thank  Dr. V. Cheianov and Prof. A. J. Niemi and for discussions
and  Dr. D. A. Gorokhov   for discussions and communicating the references \cite{kho,Bla}.

\end{multicols}

\begin{thebibliography}{999}

\bibitem{aaa}
A.A. Abrikosov Sov. Phys. JETP {\bf 5} 1174 (1957)

\bibitem{hind}
For a review see
M.B. Hindmarsh,   T.W.B. Kibble  Rept.Prog.Phys. {\bf  58} 477 (1995)

\bibitem{volo}
for a brief review see e.g. G.E. Volovik 
{\it
Proccedings of International Symposium on Quantum Fluids and
    Solids (QFS 2000)}, Minnesota, June 6-11 (2000);
preprint cond-mat/0005431 

\bibitem{Doniach}
For a recent work and citations see
D. Das, S. Doniach Phys. Rev. B {\bf 60}, 1261 (1999)


\bibitem{sc}
For a review see e.g.
M. Randeria. in: Bose-Einstein Condensation,
edited by A. Griffin, D. W. Snoke., and S. Stringary.
New York, Cambridge University Press, 1995. p.355-392.
R. Micnas, J. Ranninger and S. Robaszkiewicz  Rev. Mod. Phys. {\bf 62}, 113 (1990)
E.Babaev and H.Kleinert  Phys. Rev. B {\bf 59} 12083 (1999) 
(preprint cond-mat/9907138)
E. Babaev  Int. J. Mod. Phys. {\bf A} 16 1175 (2001) (preprint hep-th/9909052). 
and numerous references therein.



\bibitem{fetter}
For a general discussion see e.g. A. L. Fetter and J. D. Walecka
{\it Quantum theory of Many-Particle Systems}, McGraw-Hill, New York 
(1971)

\bibitem{comm} In this context we should remark that even a
 vortex in a BCS
superconductor  accumulates electric charge in the core
although the mechanism is different  \cite{kho,Bla}.


\bibitem{kho}
D.I. Khomskii and A. Freimuth, Phys. Rev. Lett. {\bf 75} 1384 (1995);

\bibitem{Bla}
G. Blatter, M. Feigel'man, V. Geshkenbein, A. Larkin, A. van
Otterlo Phys. Rev. Lett. {\bf 77}, 566 (1996); M.V. Feigel'man
V. B. Geshkenbein, A. I. Larkin, V. M. Vinokur
Pis'ma Zh. Eksp. Teor. Fiz. {\bf 75} 1384 (1995) [JETP Lett. {\bf 62}
834 (1995)]; A. van Otterlo, M. Feigel'man, V. Geshkenbein, and G. Blatter
 Phys. Rev. Lett. {\bf 75}, 3736 (1995).


\bibitem{antti}L.D. Faddeev and A.J. Niemi
Phys. Rev. Lett. {\bf 85}, 3416 (2000)
L.D. Faddeev, L. Freyhult, A. J. Niemi, P. Rajan
preprint physics/0009061.



\end{thebibliography}
\end{document}